*Original Article*

# A Remote Baby Surveillance System with RFID and GPS Tracking

Ruven A/L Sundarajoo[1], Chung Gwo Chin[2], Pang Wai Leong[3], Tan Soo Fun[4]

*[1,2,3] Faculty of Engineering, Multimedia University, Malaysia.*
*[4] Preparatory Centre for Science and Technology, University Malaysia Sabah, Malaysia.*

*[2]Corresponding Author : gcchung@mmu.edu.my*

***Abstract -*** *In the 21ˢᵗ century, sending babies or children to daycare centres has become more and more common among young guardians. The balance between full-time work and child care is increasingly challenging nowadays. In Malaysia, thousands of child abuse cases have been reported from babysitting centres every year, which indeed triggers the anxiety and stress of the guardians. Hence, this paper proposes to construct a remote baby surveillance system with radio-frequency identification (RFID) and global positioning system (GPS) tracking. With the incorporation of the internet of things (IoT), a sensor-based microcontroller is used to detect the conditions of the baby as well as the surrounding environment and then display the real-time data as well as notifications to alert the guardians via a mobile application. These conditions include the crying and waking of the baby, as well as temperature, the mattress's wetness, and moving objects around the baby. In addition, RFID and GPS location tracking are implemented to ensure the safety of the baby, while white noise is used to increase the comfort of the baby. In the end, a prototype has been successfully developed for functionality and reliability testing. Several experiments have been conducted to measure the efficiency of the mattress's wetness detection, the RFID transmission range, the frequency spectrum of white noise, and also the output power of the solar panel. The proposed system is expected to assist guardians in ensuring the safety and comfort of their babies remotely as well as prevent any occurrence of child abuse.*

***Keywords -*** *Child abuse, Global positioning system, Internet of things, Radio-frequency identification, White noise.*

## 1. Introduction

In this new century, the majority of guardians are doing their daily jobs, which require most of their time in the office. Hence, guardians will be considering sending their babies or children to the babysitting centre or hiring a caregiver for non-stop nursing of babies when they are busy at home or work. However, these two methods may not be convenient for parents according to their demands. Most importantly, guardians do not get certainty about their babies' safety in both of the cases, and a lot of news has been reported worldwide about abused children [1, 2] as well as babies' casualties due to careless mistakes made by the caregivers, as filed in Malaysia [3, 4]. For example, nearly 1,000 children are reported as victims of child abuse and neglect in Malaysia each year, based on the statistics given by Kementerian Kesihatan Malaysia (KKM) [5]. Therefore, children's well-being has recently become a hot research topic in the field of health and social sciences [6].

According to [7], there are different types of child abuse, such as physical abuse, emotional abuse, sexual abuse, neglect, and domestic violence. It may have an impact on their physical and mental health, as well as their education, which may result in behavioural consequences, including an increased risk of schizophrenia and psychosis throughout their lives, as reported in [8]. It is also found in [9] that child abuse, especially happening to those children with disabilities, is one of the contributors to the high school dropout rate in some countries. In addition, based on the research in [10], abuse as well as severe or fatal injuries were more likely when a male caregiver was present. In contrast, abuse was substantially less likely when a female caregiver was present, except for a female babysitter.

On the other hand, recent research [11, 12] indicates a vital increase in child abuse cases around the world during the Coronavirus Disease 19 (COVID-19) pandemic. One of the reasons behind this phenomenon is that the children have to be supervised by others whenever their parents need to undergo self-quarantine practice for a period of 7 to 14 days due to COVID-19. Thus, it is crucial to preserve the children's well-being and protect them from being a victim of child abuse from time to time. However, what is a better solution to get rid of the stress and anxiety of guardianship to protect their babies or children against potential dangers even without scarifying the parents' working time or spending more money to obtain better nursing services?

From this perspective, a contemporary baby care system could be one of the solutions. With the advance of the latest technologies, a monitoring system is an automatic alert system that can detect the baby's condition either in typical or perilous circumstances. As reported in [30], this solution can be implemented even before the baby is born. Multiple integrated modules such as mother care, baby care, doctor consultation, etc. were used to assist the mother during pregnancy time. Meanwhile, a paper in [14] proposed a monitoring system for babies in the incubator.

The paper provides a solution to the abovementioned problem by alerting the neonatal nurses so that preventive measures can be taken to avoid the theft being held in the incubator rooms of the baby as well as to measure parameters such as heartbeat, temperature, blood pressure, wetness, and movement of the baby.

Besides that, a baby cradle system was designed [15] to automatically turn on the cradle to assist the baby in sleeping whenever crying was detected and an alert was sent to the user through the global system for mobile communication (GSM) network. An extended system called the Internet of Things-based baby monitoring system (IoT-BBSM) was developed in [16] to include the facility of IoT, while another advanced system was designed in [17], to sense the baby's crying, moisture, and ambient temperature and share the data with their parents through an IoT platform. All of the above studies used a camera to monitor the sleeping baby.

**Table 1. Comparison of the existing and proposed systems**

| Research Work | Specification | | | | |
| | Baby's Condition | Location Tracking | Environmental Factor | Real-Time Data / Notification | Additional / Special Feature |
|---|---|---|---|---|---|
| Research paper [14] | Heartbeat, temperature, blood pressure, and movement | RFID | Mattress wetness | Data sharing & alerts using Lab view & GSM | 1. Detect theft by using heartbeat and RFID. |
| Research paper [15] | Crying | - | Mattress wetness | Only alerts using GSM | 1. Assist the baby's sleep using a cradle. 2. Using a camera. |
| Research paper [16, 17] | Crying, moisture, and ambient temperature | - | - | Data sharing & alerts using IoT | 1. Assist the baby's sleep using a smart cradle. 2. Using a web camera. |
| Research paper [18] | Temperature, crying, moisture, and movement | - | - | Only alerts using GSM | 1. Using a wireless camera. |
| Research paper [19] | Pulse rate, temperature, humidity, CO2, and weight | | Mattress wetness | Data sharing & alerts using IoT (Thinkspeak) and mobile app | - |
| Research paper [20] | - | - | Temperature | Data sharing using the web browser | 1. Identify potential treatments in the room 2. Using a Raspberry camera. |
| Research paper [21] | Movement | - | - | Data processing using Embedded Vision in IoT | 1. Detect the behaviour of the baby through motions. 2. Using a Raspberry camera. |
| Research paper [22] | Crying and temperature | - | Temperature and mattress wetness | Data sharing & alerts using IoT (Blynk) | 1. Using lighting, lullaby play, and cooling fan. |
| Proposed system | Crying, position, and movement | RFID and GPS | Temperature, mattress wetness, and moving objects | Data sharing & alerts using IoT (Blynk) and mobile app | 1. Identify potential treatments in the room 2. Comfort the baby's sleep using white noise. 3. Using LEDs and IP cameras. 4. Can be powered by solar. |

On the other hand, the GSM network was used to send readings of the movement of the baby, body temperature, and pulse rate to the user [18]. Again, the work was expanded in [19] with additional sensing of the baby's moisture and weight, as well as a new alarm-triggering system via an IoT-based mobile application. A security system using a Raspberry Pi camera was also reported in [20] to detect any possible threat in a room, and the babies' behaviour was analyzed using a chart analysis based on their motions and heartbeats using embedded vision in IoT [21]. Finally, Blynk, an IoT platform, was used [22] to demonstrate crying detection on a mobile phone. Nonetheless, most of these projects do not include real-time safety monitoring of the baby in case immediate action needs to be taken by the guardians. Furthermore, playing white noise was presented as a more effective non-pharmacological method for the crying of colicky babies than swinging [23].

This research indicated that playing the white noise can decrease the daily crying duration as well as increase the sleeping duration of the babies compared to the swinging cradle method. In conclusion, the latest technologies such as IoT, embedded sensor systems, and mobile applications can be used to develop an advanced baby care system, but the existing work can still be improved further to ensure the safety and comfort of the babies. In this paper, a remote baby surveillance system with radio-frequency identification (RFID) and global positioning system (GPS) technologies [24] is proposed. The developed prototype will be sensor-assisted to collect real-time information about the baby's conditions, such as standing position, awakening, crying, and geometrical location, as well as the conditions of the surrounding environment, such as temperature, mattress wetness, and moving objects around the baby bot. With the incorporation of IoT, a wire/WIFI platform is set up to interconnect all the input/output units through a microcontroller and to update real-time data and notifications to the user's smart devices at no cost. An RFID and GPS tracking module are included in this design to assist in locating the baby or child within a confined area to avoid any unpredictable incident before it happens to them. In addition, a white noise machine is also used to produce music that will able to comfort the sleeping baby.

The comparison of the existing and proposed systems has been summarised in Table 1. None of the existing works provides comprehensive location tracking for baby safety. Although some of the studies propose to monitor health conditions such as heartbeat, blood pressure, and others, these sensors are of higher cost and will not be considered in the proposed model since this paper aims to develop a low-cost prototype for domestic usage. On the other hand, environmental factors such as the detection of moving objects are added to improve the security and safety of the baby as compared to the existing work. Furthermore, the proposed system also provides an additional feature for comforting the baby's sleep using white noise, and the developed prototype can be powered by either electrical or solar energy.

This paper has the following organization. In Section 2, the detailed model of the proposed hardware and software system is described. Then, the demonstration of the hardware prototype and software implementation is presented in Section 3. Last but not least, the contributions of this paper as well as future recommendations are summarised.

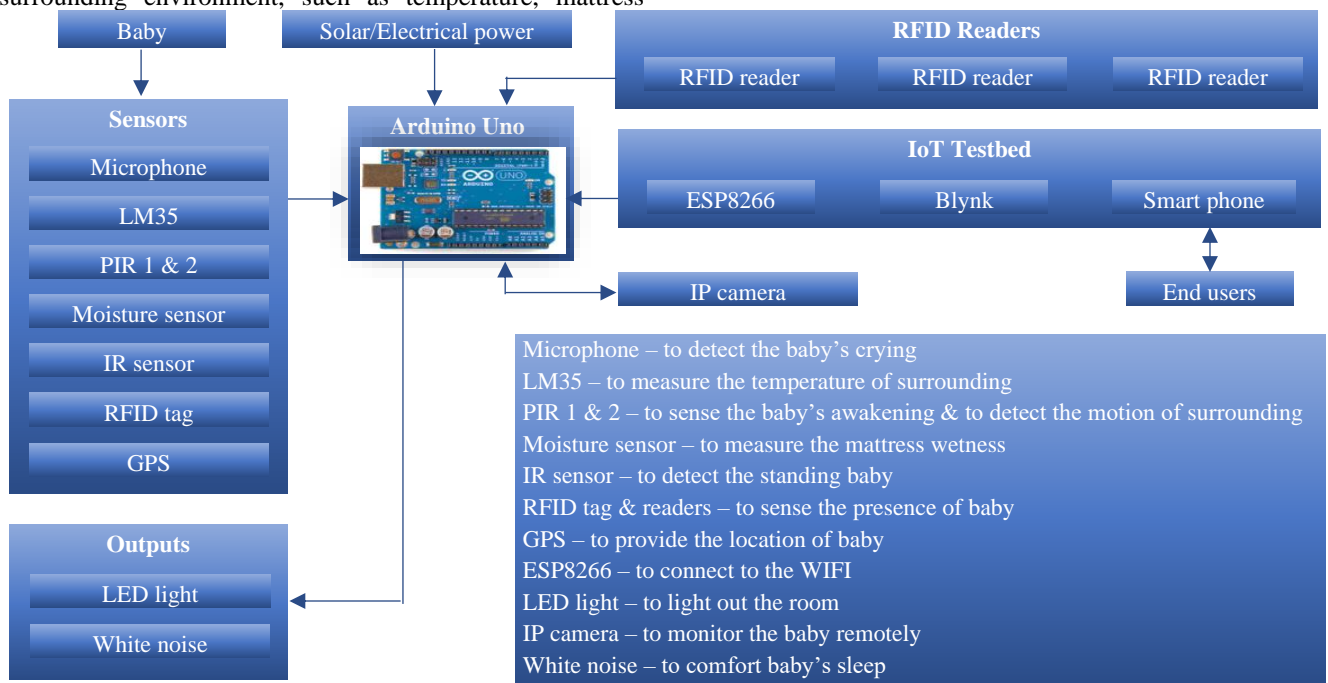

**Fig. 1 Block diagram of the proposed system**

## 2. System Model

### 2.1. Hardware Description

Figure 1 shows the block diagram of the proposed system. The system model in this project is designed to assist guardians to ensure their baby's safety and comfort and also remotely track their baby's location using the latest technologies such as RFID and GPS. The following is a description of the system model feature:

#### 2.1.1. Sensor-assisted Microcontroller Unit

Various sensory devices are either attached to the baby itself or embedded in the baby cot to obtain real-time data about the conditions of the baby as well as the surrounding environment in a confined area. These include a microphone to identify the baby's crying; an infrared (IR) sensor to detect if the baby is standing in an unsafe position; a passive infrared (PIR) sensor to detect the baby's awakening; and a moisture sensor to measure the mattress's wetness in the baby cot. For the surrounding environment, sensors such as the LM35 are used to measure the temperature, while a PIR is used to detect any possible movement of an unknown object in the room.

The Arduino Uno, a microcontroller [25], is used to collect and process the data, so that appropriate control signals will be generated to the output devices, such as a light emitting diode (LED) light and an internet protocol (IP) camera, in order to allow the guardians to keep an eye on their baby if necessary. In addition, a white noise machine is introduced in this project to improve the baby's sleep, as reported in [23]. All these components obtain the required energy supply either from electrical or solar power.

#### 2.1.2. RFID Tracking Module

RFID technology is used to track the location of the baby within the room or house. The passive MFRC522 RFID, with a frequency range of 13.56 MHz, is implemented in this project since it is easy to install and has a considerably lower cost and energy consumption. However, it only has a reading range of 1 cm to 10 cm.

A wearable RFID tag is placed on the baby, and several RFID readers are installed in the house. Whenever the baby trespasses on a specific area, such as a door, an alarm and notification are triggered. This approach allows the guardians to ensure the safety of their baby 24 hours a day without the need for human inspection from time to time.

#### 2.1.3. GPS Locator

If the RFID is triggered, the guardians can trace the location of their baby using a GPS unit that is attached to the baby. This GPS will only start to locate and send the geometrical longitude and latitude of the baby to an online platform named Firebase [26] whenever the baby is not inside the house. Firebase is a database that allows users to send data to it as well as access it securely. By knowing the real-time location, an immediate response can be provided from the guardians to their baby caretakers before the occurrence of any unhappy incident.

#### 2.1.4. IoT Testbed

To remotely control the baby care system, IoT technology is used to send notifications to the guardians as well as to allow them to activate the basic output functions of the system using any smartphone. Through this testbed, all the input and output (I/O) devices are linked together using a cable or Wi-Fi internet connection. In this research, an online IoT platform named Blynk is selected due to its compatibility with both iOS and Android devices [27]. The platform can update and display any notifications generated by Arduino on mobile phones.

### 2.2. Software Description

The flow chart of the proposed system is shown in Figure 2. After the prototype is successfully connected to the local Wi-Fi internet, all the sensors used in this project will be activated, and the Arduino Uno will start to receive the input signals sent from each individual sensor and RFID reader.

Firstly, the system will connect to the WiFi internet and turn on every sensor used in the project. Then, it will start to check if the RFID tag worn on the baby is detected. If so, it will extract the values of the latitude and longitude from the Firebase database generated from the GPS and send a notification to the smartphone via Blynk.

Secondly, the system will proceed to check if the baby is crying or awake. Then counters will be increased to store the number of occurrences and send this information with alerts to the smartphone.

Lastly, the system will also check if any moving objects are detected around the baby cot, if the room temperature is too high or too low, and if the mattress of the bed is too wet. As long as any motion has been detected, the temperature reading is more than 24 °C or less than 20 °C and the moisture level's reading exceeds 20%, it will automatically send alerts to notify the guardians through the smartphone.

In the proposed system, the alert's triggered condition for the temperature is set at 20 °C to 24 °C since this is the comfortable range for the babies. It also provides early warning of any unwanted incidents, such as the occurrence of a fire around the baby. Nevertheless, the alert's triggered condition for the moisture level is chosen to be more than 20% based on the testing results obtained in a simple experiment, which will be illustrated further in the next section.

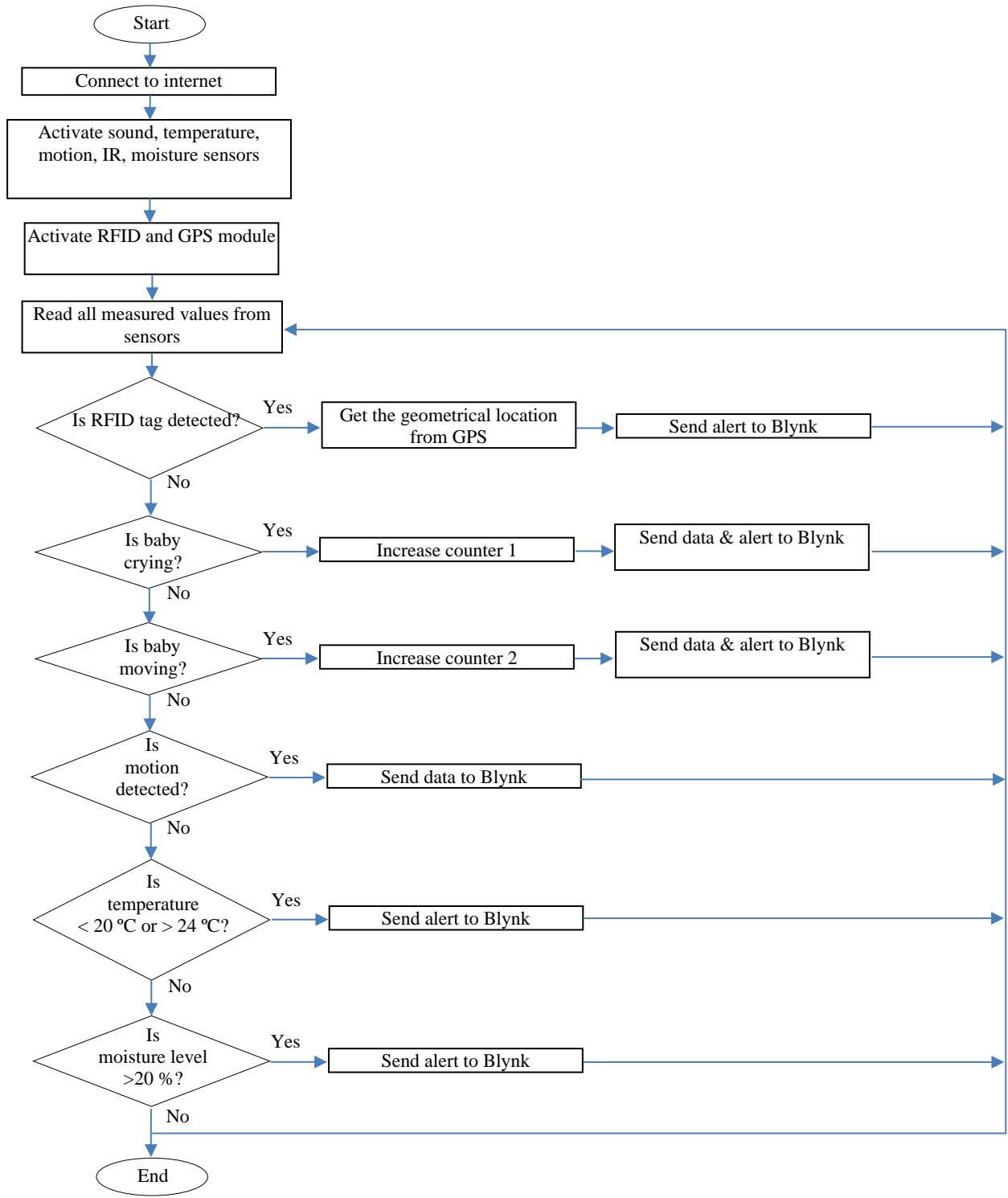

**Fig. 2 Flow chart of the proposed system**

# 3. Result Demonstration and Discussion

## 3.1. Hardware Prototype

The prototype of the proposed system has been successfully constructed, as shown in Figure 3. The microcontroller unit has been packaged in a box with all the sensors, wireless modules, and output components attached externally. The baby cot is equipped with sound, IR, motion, temperature, and moisture sensors, as well as RFID and GPS modules.

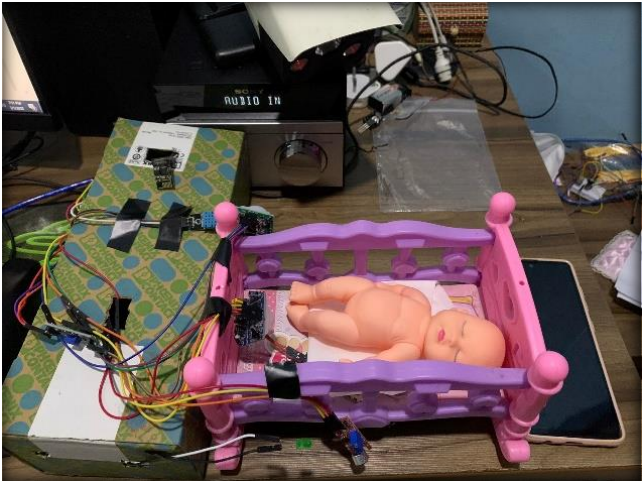

**Fig. 3 Hardware prototype**

If the baby cries or makes any loud noise, the system will detect and count the number of occurrences. If the baby awakes, moves around, and stands on the bed, the system will notify the guardians. Meanwhile, the second motion detector around the baby's cot will also sense any moving entities in the room, ensuring the baby's safety against any intruders or animals. The temperature of the surroundings is also measured to take the necessary action to maintain a comfortable environment. All of these detections will trigger the notifications or measured values and display them on the smartphone.

Besides that, some of the components can be controlled remotely by the guardians, such as the LED light and IP camera, as shown in Figure 3. They can monitor their baby at any time whenever the smartphone shows any alerts or notifications such as the baby's awakening, motion detection, hot weather, etc. Additional components such as cooling fans, alarms, etc. can also be easily added to the system if needed.

### 3.1.1. Temperature Measurement

Several experiments have been carried out to test the accuracy of the temperature sensor under various environmental conditions, such as cold (less than 20 ºC), normal (around 22 ºC), and hot (more than 24 ºC). A cold environment is created by turning on the air con with a temperature setting of 17 degrees, while a hot condition is created by turning off all the cooling facilities in the room such as the air-con and fan. All temperature measurement is taken in the morning before the room temperature is affected by the hot sun after 12 pm.

Table 2 shows the comparison of the LM35 temperature sensor's readings and the thermometer's readings. Each reading is taken from the average of five measured values to minimize the effect of human errors.

Given that $X$ is the temperature sensor's reading and $Y$ is the thermometer's reading, the relative error, $e_R$ is calculated as:

$$e_R = \frac{|X-Y|}{Y} \times 100\% \qquad (1)$$

The results obtained from Table 2 have relative errors of between 1% and 10%, which is not accurate enough. Hence, a minor adjustment has been introduced in the coding by adding a constant offset value to the sensor's input reading to improve the accuracy of the LM35 temperature sensor. Table 3 shows the second comparison of the LM35 temperature sensor's readings, $X$, and the thermometer's readings, $Y$, after the modification. The relative error of the measured temperature between the LM35 temperature sensor and the thermometer has been successfully reduced to below 5% only.

**Table 2. Comparison of the LM35 temperature sensor's readings and the thermometer's readings before adjustment**

| LM35 temperature sensor's readings (ºC) | Thermometer's readings (ºC) | Relative error (%) |
|---|---|---|
| 17.5 | 18.7 | 6.42 |
| 17.4 | 19.3 | 9.84 |
| 20.5 | 22.1 | 7.24 |
| 22.5 | 22.9 | 1.75 |
| 25.2 | 26.1 | 3.45 |
| 24.3 | 26.6 | 8.65 |

**Table 3. Comparison of the LM35 temperature sensor's readings and the thermometer's readings after the adjustment**

| LM35 temperature sensor's readings (ºC) | Thermometer's readings (ºC) | Relative error (%) |
|---|---|---|
| 18 | 18.9 | 4.76 |
| 18.5 | 19.1 | 3.14 |
| 22.2 | 22.4 | 0.89 |
| 22.7 | 23.2 | 2.16 |
| 25.7 | 26.1 | 1.53 |
| 25.3 | 26.5 | 4.53 |

### 3.1.2. Mattress's Wetness Measurement

For the wetness measurement, Figure 4 shows the comparison of the moisture sensor readings for two different types of baby mattresses: polyester fibre (Mattress A) and cotton fibre (Mattress B). They are the two common types of material used for baby mattresses on the market. The sensor is attached to the mattress in the baby cot and its measurement is taken by pouring different amounts of water onto the mattress for a time interval of 30 s to allow the mattress to absorb the water as well as the reading to reach its stability.

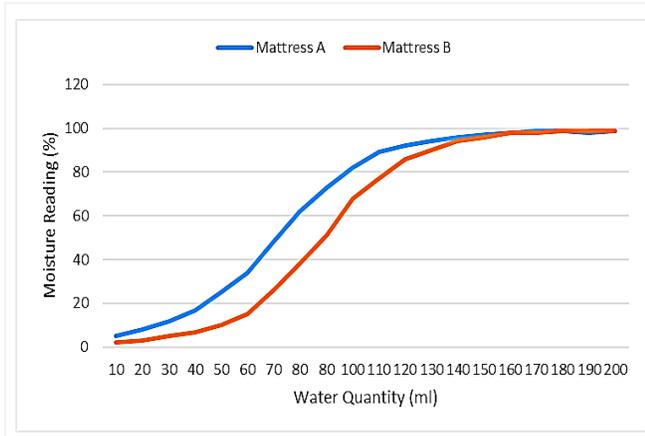

**Fig. 4 Comparison of the Moisture Sensor's Readings**

Notice that the sensor has a reading of less than 20% when the amount of pouring water is less than 50 ml for polyester and 70 for cotton, respectively. However, when the mattress is unable to absorb the amount of water poured on it, it begins to rise quickly at 40 ml for polyester and 60 ml for cotton and reaches nearly 100% at 160 ml for both polyester and cotton. It is clear that cotton has better water absorption as compared to polyester fibre. In this project, a notification will be triggered to the guardian whenever the reading exceeds 20% since the baby will start to feel uncomfortable if the mattress cannot absorb the water effectively.

### 3.1.3. RFID Tracking

For the RFID module, Figure 5 shows that the baby is wearing an MFRC522 RFID tag and the RFID reader is installed on a door at an approximate height of 3 cm from the ground. Whenever the baby trespasses the door, the RFID reader will detect the signal from the tag and send an alert to the guardians. This is to make sure that the baby stays within a confined area without the presence of the babysitter. It can be implemented for all the areas in the house by just installing RFID tags on each exit and entrance. However, the RFID detection is only limited to a perpendicular distance of up to 6 cm if the baby is crawling and 1 cm if the baby is standing, as shown in Table 4 since the passive RFID tags with a frequency range of 13.56 MHz are used in this project due to budget constraints. This distance can be extended by using higher-frequency RFID technology.

**Table 4. Detection range of RFID**

| Perpendicular Distance (cm) | Crawling Baby | Standing Baby |
|---|---|---|
| 1 | Yes | Yes |
| 2 | Yes | No |
| 3 | Yes | No |
| 4 | Yes | No |
| 6 | Yes | No |
| 8 | No | No |

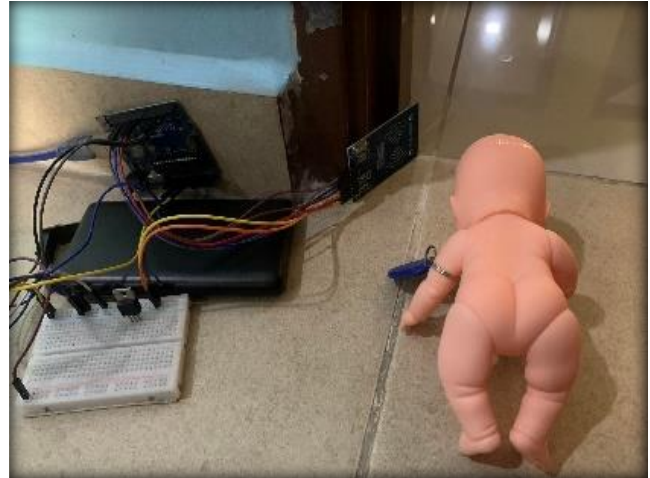

**Fig. 5 Wearable RFID tag, and RFID reader**

### 3.1.4. White Noise Testing

There is also a white noise machine placed near the baby cot, as shown in Figure 6. It is used to generate white noises, which can reduce unwanted noises as well as provide a better and healthier sleeping environment for the baby [23].

Figure 7 depicts the frequency spectrums of the machine's white noise and the unwanted noise produced by the construction (drilling). The spectrum is plotted using Friture [28], a free online tool that can capture surrounding sound and present its frequency spectrum instantly. It is observed that the white noise occupies a flat spectrum overlapping a wide range of frequencies (from 250 Hertz to 5500 Hertz), which includes the construction noise. According to the principle of superposition theorem [29], white noise is able to flatten the spectrum of the other noises and hence minimize the noise effect brought by the unwanted noises.

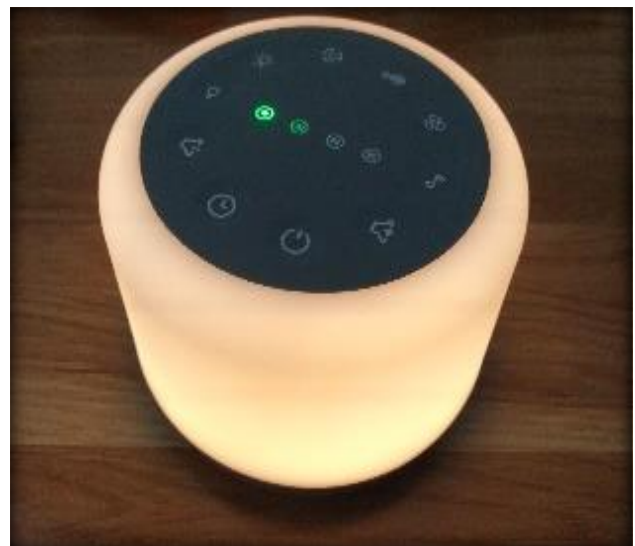

**Fig. 6 White noise machine**

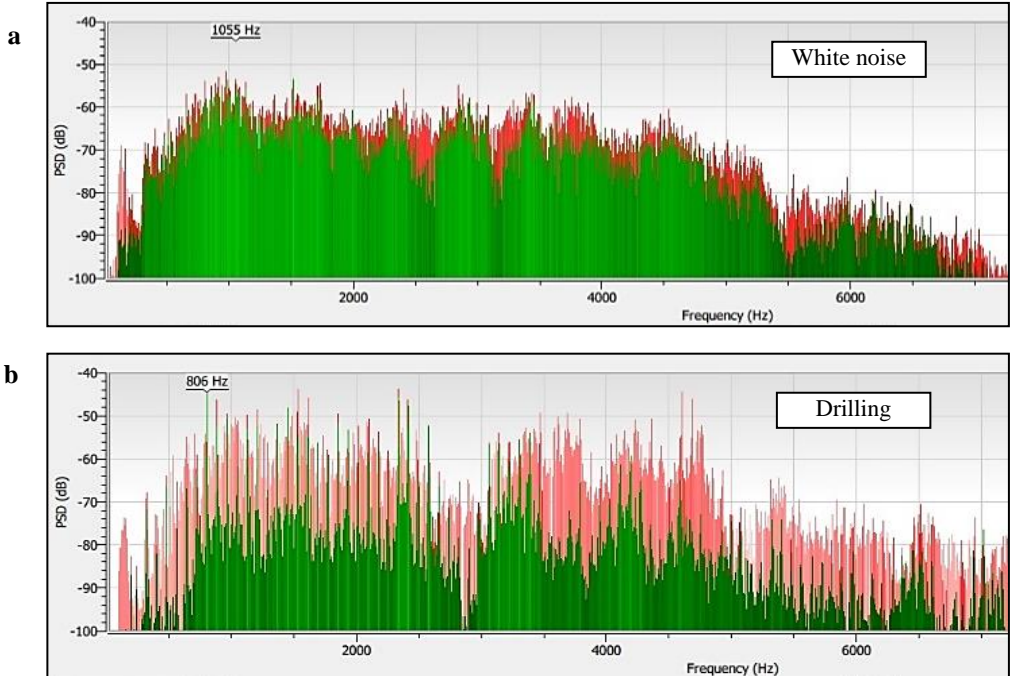

**Fig. 7 Frequency spectrum of (a) white noise, and (b) drilling noise**

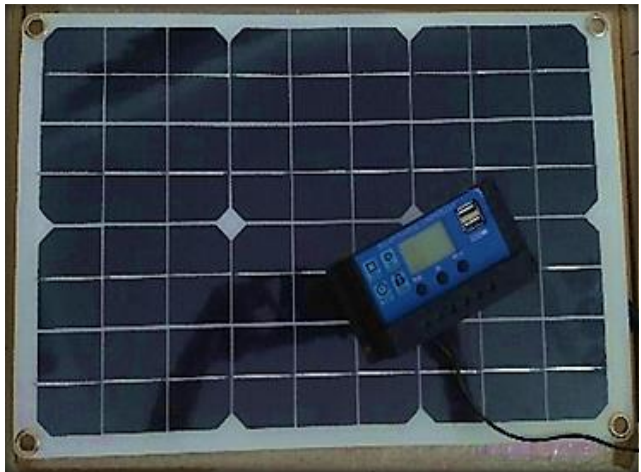

**Fig. 8 Solar panel and controller**

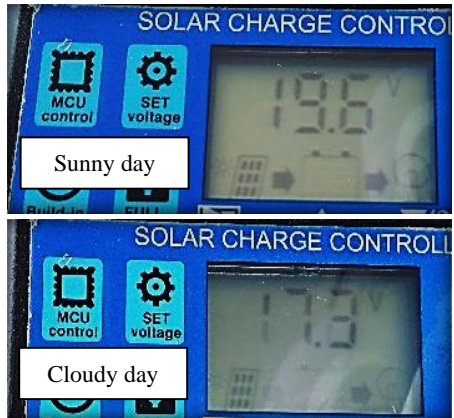

**Fig. 9 Readings of the controller**

### 3.1.5. Solar Power Measurement

Last but not least, the solar panel and its controller are shown in Figure 8. The solar panel generates an average voltage reading of around 19.6 V and 17.3 V with and without a direct hot sun (under the cloud), as shown in Figure 9. The solar panel is placed perpendicularly under the sun to maximize its exposure to the sun's light without any obstacle in between. Research has been initiated to study the output stability of the solar panel as shown in Table 5. Each reading is taken from the average of five measured values from five different days to minimize the effect of human errors and weather. From 10 a.m. to 6 p.m., aside from when it is raining heavily, the variation of the output voltage does not exceed 17% of its maximum values recorded under the direct hot sun. Based on the obtained results, as long as it is a bright sunny day, the solar panel can generate sufficient power for the developed prototype to function properly during the daytime from 10 a.m. to 6 p.m. However, solar power cannot be used for power generation during the time from 8 p.m. to 6 a.m. without a rechargeable battery due to budget constraints.

**Table 5. Readings of the controller**

| Time | Weather Condition | | |
|---|---|---|---|
| | Direct hot sun | Cloudy (Little) | Rain (Heavy) |
| 8 a.m. | 17.7 V | 15.3 V | 0 V |
| 10 a.m. | 19.2 V | 16.8 V | 0 V |
| 12 p.m. | 19.6 V | 17.3 V | 0 V |
| 2 p.m. | 19.6 V | 17.2 V | 0 V |
| 4 p.m. | 19.4 V | 17 V | 0 V |
| 6 p.m. | 18.9 V | 16.4 V | 0 V |
| 8 p.m. – 6 a.m. | 0 V | 0 V | 0 V |

### 3.2. Software Implementation
#### 3.2.1. Data Display

Whenever the sensors used in the hardware prototype detect any changes in the conditions of the baby and the surrounding environment, the microcontroller will generate the necessary data and notifications and send them to the smartphones via the IoT platform as mentioned before. Figure 10 shows the mobile application, which displays the detection of motion and sound as indicated by "YES" or "NO". If the sensor detects any loud noises, such as a baby crying, the application will display a "YES" for sound detection and a "YES" for motion detection if any object moves around the bed or cot. This feature can immediately alert the guardians if there is any danger encountered by the baby.

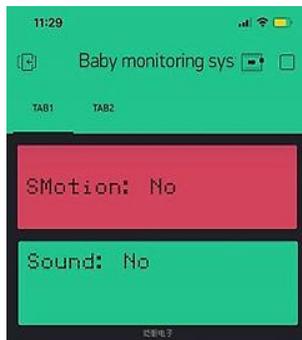

**Fig. 10 Display of motion and sound detection**

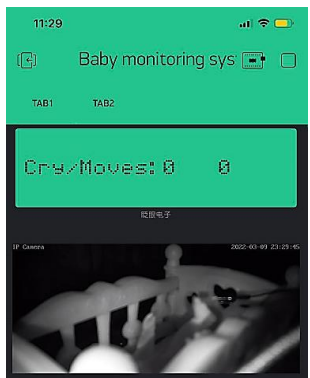

**Fig. 11 Display of crying and movement counter, and live video from IP camera**

On the other hand, the prototype will also sense any movement of the baby on the bed and store the counted number of crying and movement as displayed in Figure 11. This counting data is essential to allow the guardians to understand the behaviour of their baby throughout the day and prepare any necessary plans in the future to reduce the occurrence of the baby's crying and awakening. If needed, the guardians can also remotely turn on the view of the IP camera as shown in Figure 11 to observe the real-time conditions of the baby. This feature can also assist the guardians to make sure that the babysitting is being done properly throughout the day.

#### 3.2.2. Mobile Notifications

The mobile application is also designed to receive notifications generated by the system for the baby's crying, baby's moving, room temperature, mattress's wetness, and RFID tracking, as presented in Figure 12. They will only be triggered when the room has a temperature of more than 24 ℃ or less than 20 ℃, the baby is crying or moving, the moisture level reading of the mattress exceeds 20% and the RFID is detected. Otherwise, the user needs to check on the sensors' readings for real-time information. All these notifications alert the guardians to respond immediately to their babysitter for crucial checks as well as necessary actions are taken on their baby and its surrounding environment.

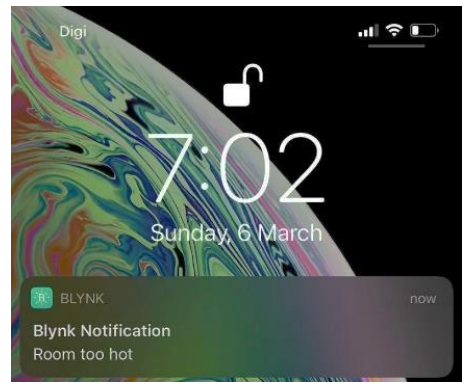

**a**

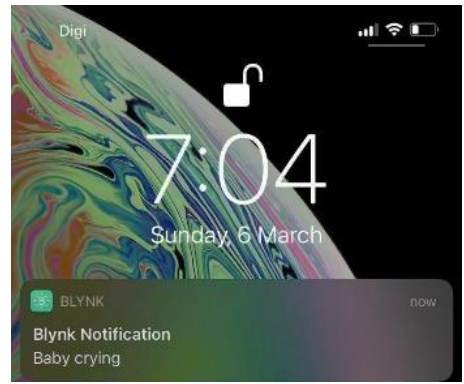

**b**

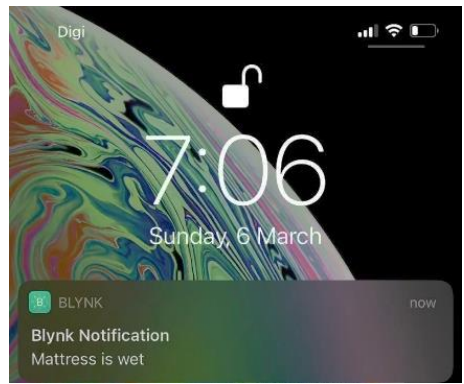

**c**

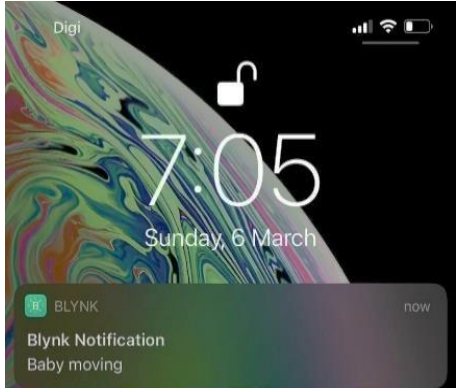

**d**

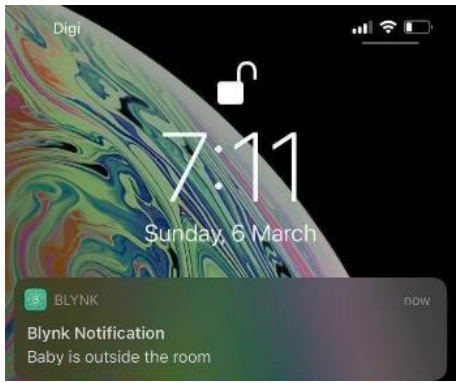

**e**

**Fig. 12 Notifications on (a) room temperature, (b) baby's crying, (c) mattress wetness, (d) baby's moving, and (e) RFID tracking**

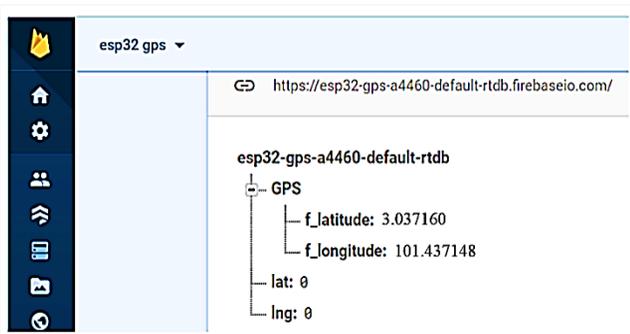

**Fig. 13 GPS location shown in Firebase database**

### 3.2.3. GPS Location

Finally, if the RFID is triggered, the guardians can obtain the actual geometrical location of their baby through the Firebase database using smartphones or computers, as shown in Figure 13. It stores the values of the latitude and longitude obtained from the GPS unit and displays them on the website. From the values of latitude and longitude shown in Figure 13, it indicates that the baby is located around Klang, Malaysia. The value is sent to Firebase using the API key with the hostname and the token. The API key for Firebase authentication can be found on their website [26].

### 3.3. Final Achievement

In conclusion, various practical experiments and prototype demonstrations have been conducted and presented to improve the performance accuracy, robustness, and consistency of the developed system compared to the existing work. For instance, the temperature measurement has recorded less than 5% of relative error after some calibrations have been done on the sensor. The effective detection ranges of the RFID and also the actual water absorption rate of the moisture sensor have been found through several testing too.

This is to make sure that the system generates the required notifications to the users based on the correct detection of the sensor. Moreover, the white noise machine and solar panel are tested for their functionality, although further research is needed to verify the effectiveness of white noise on the sleeping baby. As compared to the previous existing works, the proposed work in this paper also offers a more complete solution for the modern baby care system with the implementation of RFID and GPS tracking modules to ensure the safety of the baby by tracking the location of the baby 24-hours. All these features can be applied effortlessly using the online IoT platform and mobile application.

## 4. Conclusion

New centuries of guardians do not ensure their babies' safety whenever they send them to the babysitting centre. Due to the careless mistakes made by the caretaker, there have been a lot of cases of child abuse or baby casualties in Malaysia recently. In order to overcome this problem, a prototype has been successfully constructed in this project to provide a remote baby surveillance system to the guardians, which allows them to closely observe their baby's conditions via a mobile application.

Using advanced technologies such as RFID and GPS, guardians can ensure the safety and comfort of their babies 24 hours a day by receiving alerts and notifications via the IoT platform, which indicate that the baby is crying, possibly due to being scared by any objects around the baby cot, or that the room conditions are not comfortable, or that the baby is located in a dangerous area alone.

The guardians can take immediate action by turning on the LED light and observing their baby on the live video. They can also contact the caregiver to deploy the white noise music to comfort the baby's sleep. Hence, the proposed system allows the monitoring and controlling of different processes remotely and efficiently, which provides an opportunity for guardians to have better child care for their children.

From future perspectives, higher frequency RFID technologies such as Near-Field Communication (NFC) can be implemented to enhance the functionality of the system, and, artificial intelligence (AI) can be applied to perform necessary actions immediately and automatically without the need for human intervention.

The prototype can also be extended with additional sensors such as smoke or flame detector, humidity sensor, etc., and output units such as alarms, cooling fans, etc. to enhance the functionality of the system without increasing the cost significantly.

## Funding Statement

This research project is fully sponsored by Internal Research Fund (MMUI/210076), Multimedia University.

## Acknowledgments

The authors confirm their contribution to the paper as follows: Study conception and design: Author 2, Author 4; Prototype development and data collection: Author 1; Analysis and interpretation of results: Author 3; Draft manuscript preparation: Author 2. All authors reviewed the results and approved the final version of the manuscript.